\documentstyle[preprint,tighten,aps,prd]{revtex}

\begin{document}

\preprint{\parbox{\hsize}{\flushright FAU-TP3-98/5\\hep-th/9803177\\\ }}
\title{Structure and Dynamics of  Monopoles in Axial-Gauge QCD%
  \footnote{Copyright 1998 by The American Physical Society.}}
\author{O.~Jahn and F.~Lenz}
\address{Institut f\"ur Theoretische Physik III \\
  Universit\"at Erlangen--N\"urnberg \\
  Staudtstra{\ss}e 7 \\
  D-91058 Erlangen, Germany}
\date{24 March 1998}
\maketitle
\begin{abstract}
  An investigation of singular fields emerging in the process of transforming
  QCD to the axial gauge is presented. The structure of the singularities is
  analyzed. It is shown that apart from well known neutral magnetic monopole
  singularities, the field configurations also exhibit singularities in their
  charged and transverse components. This complex singularity structure
  guarantees finite non-Abelian field strength and thus finite action if
  expressed in terms of gauge fixed fields.  The magnetic monopoles are
  characterized by two charges which are shown to determine the Pontrjagin
  index.  Qualitative dynamical aspects of the role of the monopoles are
  discussed. It is argued that the entropy associated with monopoles increases
  with decreasing temperature and that the coupling to quantum fluctuations
  favors monopole-antimonopole binding.
\end{abstract}

\narrowtext

\section{Introduction}

A fundamental and possibly far reaching difference in the formal structure of
QED and QCD emerges when formulating these gauge theories in terms of physical,
unconstrained variables. While this can be achieved in QED by a global -- for
all field configurations valid -- choice of variables, this is possible in QCD
only at the expense of introducing coordinate singularities and thus of
including singular gauge field configurations in such gauge fixed formulations
\cite{SING78}. It is tempting to associate the vastly different physical
properties of these two theories with this basic geometrical difference.
Formation of Gribov horizons \cite{GRIB78} or condensation of magnetic
monopoles \cite{MAND80,THOO81} represent two prominent attempts in which the
emergence of confinement is linked directly to the presence of singularities in
gauge fields after gauge fixing. It appears that QCD lattice calculations
\cite{KRSW87,POLI97} confirm the dynamical picture of confinement via the dual
Meissner effect.  Conclusive results have been obtained within the ``maximal
Abelian gauge''. It is, however, not unambiguously established that these
mechanisms are operative also in other formulations of Abelian projected QCD,
and the relation to mechanisms for confinement in fundamentally different
gauges, such as the Coulomb gauge, is not understood.  Lattice calculations
also seem to support the hypothesis of ``Abelian dominance'' \cite{EZIW82}
according to which long-range phenomena and specifically confinement are
dominated by the ``neutral'' components of the gauge fields. For instance,
contributions of the charged components to the string tension seem to be
negligible \cite{SUYO90}. While favored in lattice calculations, the maximal
Abelian gauge is not particularly suited for explicit analytical
investigations. Although selected analytical results have been obtained, in
particular concerning the connection between monopoles and instantons
\cite{CHGU95,BROT97}, a complete resolution of the maximal Abelian gauge
condition is not possible nor can the singular field configurations be
classified systematically.  In this work we will present an analytical study of
the structure and the dynamical role of singular field configurations within
the modified axial (temporal) gauge representation of QCD. The reason for this
particular choice is primarily technical; it is the only case where complete
gauge fixing can be explicitly performed and by which a representation of QCD
can be achieved in terms of unconstrained variables only. The resulting
formulation has remarkable properties which are useful for clarification of
both the formal structure and the dynamical implications of singular field
configurations. The crucial step in the gauge fixing procedure at which
singularities are generated is the diagonalization of the (untraced) Polyakov
loop. The trace thereof is the order parameter characterizing the confined and
deconfined phases and its values determine the presence of singularities.
Furthermore, in this gauge, this order parameter is represented by an
elementary rather than a composite field.
 
Gauge fixing by which coordinate singularities are generated is achieved
through gauge transformations. Therefore, gauge invariant quantities associated
with fields which become singular by gauge fixing must remain finite. This
determines to a large extent the basic structure of singular field
configurations. In particular singularities in the gauge field must not show up
in the (non-Abelian) field strength and thus singularities necessarily occur in
neutral and charged components of the gauge field simultaneously.  Abelian
dominance is not an immediate consequence of the singular structure of these
field configurations.  This elementary discussion is carried out in detail in
Sec.~\ref{singularities} for gauge choices which involve diagonalization of
the Polyakov loop.
It is shown that in these gauges the monopoles carry, apart from the
magnetic charge, an additional charge.  This charge assumes values in the
center of the gauge group and depends on the eigenvalues of the Polyakov loop.
With gauge invariant quantities not affected by the
presence of singularities in the gauge fields it is possible to remove those
parts of space time where the gauge fields become singular and to account for
the singularities by appropriate boundary conditions -- a procedure similar in
spirit to the Yang-Wu construction \cite{WUYA76}.  This program is explicitly
carried out in Sec.~\ref{fields} for axial gauge QCD and, as a particular
consequence, the Dirac quantization condition for charged fields moving in the
background of singular fields is obtained.  In axial gauge QCD, variables
characterizing the orientation of the Polyakov loops have been eliminated.
Therefore the singularities must be determined completely by the gauge
invariant eigenvalues of the Polyakov loops.  Explicit expressions in terms of
these eigenvalues for the gauge fixed fields close to the singularities are
given in Sec.~\ref{parameters}. As a byproduct, this discussion will also
show that the singularities have additional structures not determined by
topological arguments.  For instance singularities with continuously variable
strength may occur in transverse Abelian magnetic fields.  In general,
emergence of singularities in the gauge fixing procedure does not point to any
peculiar dynamics. However there are circumstances in which the presence of
singularities is actually required. This issue is discussed in
Sec.~\ref{instantons} where it will be shown that a non-vanishing
topological charge associated with a gauge fixed field configuration is
possible only if this configuration is singular.  With the gauge fixed fields
satisfying trivial boundary conditions, the topological charge receives
contributions exclusively from the surface of those parts of space time which
had to be removed due to the presence of singularities. In this way, a relation
between topological charge and monopole charges can be established.  (In a
formulation involving singular non-Abelian field strengths, this relation has
been given in Ref.~\cite{Reinhardt97}.)  Apart from
a short summary, the concluding section contains a qualitative discussion of
the physics content of the formalism developed up to this point.  We do not
carry out any dynamical calculation but rather interpret well known properties
of QCD in terms of monopole dynamics and address specifically the issue of
monopole condensation and its possible relevance for the properties of the
confined and deconfined phases.

\section{Singularities from Diagonalization of Polyakov Loops}
\label{singularities}

In this section we discuss the occurrence of singular gauge fields in the
process of gauge fixing. We assume that the gauge fixing procedure involves
diagonalization of the Polyakov loop as is the case in axial gauge.  The focus
of this discussion will be on the singularity structure of the fields after
gauge fixing.

Throughout this work we consider SU(2) QCD at finite temperature,
i.e., in Euclidean space with compact imaginary time direction of
length $\beta$.  The gauge potentials are assumed to be periodic up to 
a gauge transformation in time,
\begin{equation}
  \label{eq:time-b.c.}
  A_{\mu}(\beta,{\bf x}) = U({\bf x}) \left( A_{\mu}(0,{\bf x}) +
    \frac{1}{i g} \partial_{\mu} \right) U^{\dagger}({\bf x}) ,
\end{equation}
continuous before gauge fixing and to approach pure gauges at spatial
infinity.  The (untraced) Polyakov loops
\begin{equation}
  P({\bf x})= P \exp\left\{ig \int_{0}^{\beta} dx_{0}A_{0}
    \left(x_{0},{\bf x}\right)\right\} \, U({\bf x})
  \equiv \exp\left\{ig \beta a_{0}^{c}\left({\bf x} \right)  \tau^{c}\right\} 
\label{I29} 
\end{equation}  
are of particular importance.  Their definition contains the
transition function $U({\bf x})$ to account for the non-periodicity of
the gauge fields.  In this way, the eigenvalues of $P({\bf x})$ are
invariant also under non-periodic gauge transformations.  The Polyakov
loops are parametrized by the fields $a_{0}^{c}\left({\bf x} \right)$
whose modulus satisfies $0 \le a_{0}({\bf x})\le \pi/g\beta$. This restriction
does not spoil the continuity of $a_0$ as follows from the continuity of
${\rm tr}\,P=2\cos g\beta a_0$.  The restriction, however, introduces
discontinuities in derivatives similar to those appearing in the
transition from Cartesian to polar coordinates.  For the following topological
considerations, continuity is sufficient.  The discontinuities will be taken
into account in the discussion of the detailed structure of the singularities
in Sec.~\ref{fields}.  In the
absence of quarks, the expectation value of the (traced) Polyakov
loops serves as order parameter for characterization of the confined
and deconfined phases.  In the axial (temporal) gauge representation,
the Polyakov loop variables $a^{c}_{0}$ appear as elementary rather
than composite degrees of freedom. This representation of QCD is
obtained as a result of the following gauge transformation performed
in three steps \cite{LENT94},
\begin{equation}
  \Omega(x) = \Omega_{{\rm D}}\left({\bf x}\right)
  \, \exp\left\{-ig x_{0} a_{0}^{c}\left({\bf x} \right) \tau^{c}\right\}
  P \exp\left\{ig \int_{0}^{x_{0}} d t A_{0}
    \left(t,{\bf x}\right)\right\} .
\label{cs2} 
\end{equation} 
In the presence of the third factor only, the gauge transformation
would eliminate $A_{0}$ completely.  The second term reintroduces zero
mode fields, and is constructed such that the transformed gauge
fields are strictly periodic in time.  Finally $\Omega_{{\rm D}}$
diagonalizes these zero mode fields or, equivalently, the Polyakov
loops of Eq.~(\ref{I29}),
\begin{equation}
  \label{cs3}
  \Omega_{{\rm D}}\left({\bf x}\right)P({\bf x})
  \Omega_{{\rm D}}^{\dagger}\left({\bf x}\right)=
  e^{ig\beta a_{0}\left({\bf x}\right)\tau_{3}} .  
\end{equation}
As a result of the gauge fixing transformation (\ref{cs2}), the time component
of the gauge field is eliminated apart from a diagonal zero mode field,
\begin{equation}   
  \label{za1}
  \Omega\left(x\right) \left(A_{0}(x)+\frac{1}{ig}\partial_{0}\right)
  \Omega^{\dagger}\left(x\right) = a_{0}\left({\bf x}\right)\tau_{3} .
\end{equation}

The diagonalization (\ref{cs3}) can also be interpreted as a choice of
coordinates in color space. With Eq.~(\ref{cs3}), the color three direction is
chosen to be that of the Polyakov loop at given ${\bf x}$. This diagonalization
is the crucial element of the gauge fixing procedure by which singular field
configurations emerge. It is most easily studied explicitly by introducing
polar ($\theta $) and azimuthal ($\varphi $) angles in color space,
\begin{eqnarray}
  \tilde{a}^{c}\left({\bf x}\right)\tau^{c} & = &
  \tilde{{\bf{a}}}\cdot \bbox{\tau}
  = \sin\left(g\beta a_{0}^{c} \left({\bf x}\right)\tau^{c}\right) \nonumber\\
  &=& \sin\left(g\beta a_{0}\left({\bf x}\right)\right) \left(
    \begin {array}{cc} 
      \cos(\theta \left({\bf x}\right)) &
      e^{-i\varphi  \left({\bf x}\right)}\sin(\theta \left({\bf x}\right)) \\
      \noalign{\medskip}
      e^{i\varphi  \left({\bf x}\right)} \sin(\theta \left({\bf x}\right)) &
      - \cos(\theta \left({\bf x}\right))
    \end {array} \right) .
  \label{am2} 
\end{eqnarray}
Note that $\tilde{\bf{a}}$ is a continuous function of the (untraced)
Polyakov loop $P$.  With this choice of coordinates, the
matrix $\Omega_{{\rm D}}\left({\bf x}\right)$ can be represented as
\begin{equation} 
  \Omega_{{\rm D}}\left({\bf x}\right) =
  \left (\begin {array}{cc}
      e^{i\varphi  \left({\bf x}\right)} \cos(\theta \left({\bf x}\right)/2) &
      \sin(\theta \left({\bf x}\right)/2) \\\noalign{\medskip}
      - \sin(\theta \left({\bf x}\right)/2) &
      e^{-i\varphi  \left({\bf x}\right)} \cos(\theta \left({\bf x}\right)/2)
    \end {array} \right) .
\label{am3} 
\end{equation}
Consequences related to the ambiguities in the definition of $\Omega_{{\rm D}}$
will be discussed later.  To improve readability, we suppress the arguments of
$\theta$ and $\varphi$ in the following.

In general, diagonalization, or equivalently the choice of coordinates, is not
everywhere well defined and consequently (coordinate) singularities occur in
the associated transformations of the gauge fields.  Starting from an
everywhere-regular gauge field $A_{i}$, the transformed field
\begin{equation} 
  A_{i}^{\prime} \left(x\right) = \Omega_{{\rm D}}\left({\bf x}\right) 
  A_{i} \left(x\right)\Omega_{{\rm D}}^{\dagger}\left({\bf x}\right)
  + s_{i}\left({\bf x}\right)  
  \label{am5} 
\end{equation}
with 
\begin{equation} 
  s_{i}\left({\bf x}\right) = \Omega_{{\rm D}}\left({\bf x}\right) 
  \frac{1}{ig}\partial _{i}\Omega_{{\rm D}}^{\dagger}\left({\bf x}\right) 
\label{am5a} 
\end{equation}
is in general singular with $\Omega_{{\rm D}}$.  While the homogeneous term
can at most be discontinuous, the inhomogeneous term diverges in
general.  We decompose this inhomogeneous term into neutral ($\tau_3$)
and charged ($\tau_{\pm}=(\tau_1\pm i\tau_2)/2$) components with
respect to the color orientation of the Polyakov loop,
\begin{equation} 
  s_{i}\left({\bf x}\right)  =  a_{i}^{{\rm s}}\left({\bf x}\right)\tau_{3}
  + \left(\phi_{i}^{{\rm s}}\left({\bf x}\right)\tau_{+}+\text{h.c.}\right) ,
  \label{am6} 
\end{equation}
where
\begin{eqnarray}
  \label{gf14}
   a_{i}^{{\rm s}}({\bf x}) & = & -\frac{1}{2g}\left(1+\cos\theta \right)
   \partial_{i}\varphi   =
   - \frac{1+\cos\theta }{2g\sin{\theta }}
   \frac{1}{\sin{g\beta a_{0}}} \,
   \hat{\bbox{\varphi}} \cdot \partial_{i}\tilde{\bf{a}} ,
   \nonumber\\
   \phi_{i}^{{\rm s}}({\bf x}) & = & \frac{1}{2g} e^{i\varphi  }
   \left(\sin\theta \, \partial_{i}\varphi  
     + i\partial_{i}\theta  \right) =
   \frac{1}{2g\sin{g\beta a_{0}}} e^{i\varphi} \,
   (\hat{\bbox{\varphi}}+i\hat{\bbox{\theta}})
   \cdot \partial_{i}\tilde{\bf{a}}
\end{eqnarray}
with the standard choice of unit vectors 
\begin{equation}
  \label{unitvectors}
  \hat{\bbox{\varphi}} = \left(
    \begin{array}{c}
      -\sin{\varphi  }\\
      \cos{\varphi  }\\
      0
    \end{array}\right)
  \qquad \text{and} \qquad
  \hat{\bbox{\theta}} = \left(
    \begin{array}{c}
      \cos{\theta }\cos{\varphi  }\\
      \cos{\theta }\sin{\varphi  }\\
      -\sin{\theta }
    \end{array}
  \right)
  .
\end{equation}
In order to make the singularity structure manifest we have expressed $s_{i}$
in terms of the continuous field $\tilde{\bf{a}}$ present before gauge fixing
(cf.~Eq.~(\ref{am2})).  Singularities (poles) occur at points
${\bf x}^{{\rm N},{\rm S}}$,
where the Polyakov loop passes through the center of the group and does not
define a direction in color space,
\begin{equation}
z\equiv P({\bf x}^{{\rm N},{\rm S}})= \pm \openone .
\label{za2}
\end{equation}
This requirement determines a point on the group manifold $\text{S}^{3}$ and
thus, for generic cases, fixes (locally) uniquely the position
${\bf x}^{{\rm N},{\rm S}}$.  In four space the transformed gauge fields
are thus
singular on straight lines parallel to the time axis, i.e., they represent
static singular field configurations. The static nature of the singularities is
a trivial consequence of the static Polyakov loop which has been selected for
introducing coordinates in color space.  The singularities can be distinguished
by the value of the Polyakov loop and we shall refer to them as north
($g\beta a_{0}( {\bf x}^{{\rm N}}) = 0$) and south
($g\beta a_{0}( {\bf x}^{{\rm S}}) = \pi$) pole singularities according to
the respective positions of the Polyakov loop on the group manifold SU(2)
($\cong{\rm S}^3$).  Thus we can assign a ``north-south'' quantum number or
charge $z$ to this singularity (i.e., the range of $z$ is the center of the
group).  This north-south charge should not be confused with the magnetic
charge to be introduced below.  In addition to
poles, the field $a_{i}^{{\rm s}}$ also exhibits (static) string like
singularities along the line $\theta =0$ representing a surface in four space. 
The charged gluon fields too have poles at ${\bf x}^{{\rm N},{\rm S}}$ and
discontinuities along the strings $\theta=0$.  Although the points
${\bf x}^{{\rm N},{\rm S}}$ are characterized by Eq.~(\ref{za2}) in a gauge
invariant way (by the degeneracy of two gauge invariant eigenvalues), in
general those points have no particular significance in other gauges.

Before continuing with our general discussion we consider special field
configurations in which color and spatial orientations are identified,
\begin{equation} 
  \varphi \left({\bf x}\right) = \pm \varphi_{0}\equiv  \pm \arctan \frac{y}{x}
  ,\qquad
  \theta \left({\bf x}\right) = \theta_{0}\equiv\arccos \frac{z}{r}\, .
  \label{am17} 
\end{equation}
Thereby the singular field $a_{i}^{{\rm s}}({\bf x})$ in Eq.~(\ref{gf14})
becomes exactly the vector potential of a Dirac monopole \cite{DIRAC31} of
charge $\pm2\pi/g$,
\begin{equation}
  {\bf a}^{\pm} \left({\bf x}\right)  
  = \mp \frac{1}{2g}\frac{1+ \cos\theta_{0}}{r \sin\theta_{0}}
  \hat{\bbox{\varphi}}_{0} 
  \label{am18}
\end{equation}
with Abelian (neutral) magnetic field
\begin{equation}
  \label{b-Dirac}
  {\bf b}^{\pm} = {\rm curl}\, {\bf a}^{\pm} 
  =\pm \frac{1}{2g} \frac{{\bf x}}{x^3} .
\end{equation}
Associated with the singular neutral component is a singular charged component
 which is given by
\begin{equation}
  \bbox{\phi}^{\pm} \left({\bf x}\right) = \frac{1}{2gr}
  \left(\pm\hat{\bbox{\varphi}}_{0}+i\hat{\bbox{\theta}}_{0}\right)
  e^{\pm i \varphi_{0}} .
  \label{za3}
\end{equation}
 
In returning to the formal development, we introduce the  Abelian field
strength
\begin{equation}
  f_{i j} = \partial_{i}a_{j}^{{\rm s}}-\partial_{j}a_{i}^{{\rm s}}, 
  \label{za3a}
\end{equation}
an important quantity in Abelian projected theories.  It is singular at the
position of the monopoles. On the other hand, the complete non-Abelian field
strength built from the inhomogeneous term $s_{i}$ of Eq.~(\ref{am5a}) actually
vanishes,
\begin{equation}
  F_{i j}\left[s\right] =
  \partial_{i}s_{j}-\partial_{j}s_{i}+ig\left[s_{i},s_{j}\right] = 0 ,
  \label{za4}
\end{equation}
i.e., the singular Abelian field strength is exactly canceled by the
non-Abelian contribution to $F$ generated by the singularities in the
charged gluon fields.

A non-vanishing field strength can occur only at the singularities of the pure
gauge field $s_{i}$. Application of the Gauss theorem shows that the
non-Abelian field strength does not contain monopole singularities and by
application of the standard (Abelian) Yang-Wu construction \cite{WUYA76}
strings can be completely avoided.  The cancellation of the singularities in
the Abelian field strength by the non-Abelian commutator must happen quite
generally. The singular fields arise as coordinate singularities since we
insist on fixing the gauge globally.  In adopting a generalized Yang-Wu
construction with some other (non-Abelian) gauge choice on charts which cover
the neighborhood of the points ${\bf x}^{{\rm N},{\rm S}}$, occurrence of any
singularity can be avoided. It is important for the following that despite the
singularities generated in the gauge fields the gauge fixing procedure does not
affect gauge invariant quantities even at the positions of the monopoles and
along the strings. This will allow us to remove these points and strings from
space time at the expense of formulating appropriate boundary conditions for
the gauge fields.  (In a similar context, this procedure to exclude the
  singular points from space time has been discussed in
  Ref.~\cite{GRIE97}.)
A further consequence of the non-Abelian field strength being
finite is the rigid connection between neutral and charged singular components
of the gauge fields.  Abelian dominance is thus not realized at the level of
the singular field configurations. Instead of applying a Yang-Wu construction,
one may proceed by regularizing the singular field configurations
(cf.\ Ref.~\cite{Langfeld96}). 
In this way the ``coordinate'' transformation
corresponding to Eq.~(\ref{cs2}) does not necessarily remain a gauge
transformation and therefore values of gauge invariant quantities like
${\rm tr}\,(F^{2})$ do not remain invariant and may even become infinite
when removing the regulator.
In Ref.~\cite{Langfeld96} it is argued that monopole contributions to the
Abelian and non-Abelian terms in Eq.~(\ref{za4}) cancel.  In contrast to our
approach, infinite non-Abelian fields appear along the strings.

Unlike the points ${\bf x}^{{\rm N},{\rm S}}$ which are determined in a gauge
invariant way (cf.\ Eq.\ (\ref{za2})) the location of the strings is to a large
extent still arbitrary since the gauge transformation $\Omega$ of
Eq.~(\ref{cs2}) does not fix the gauge completely. An Abelian rotation with an
${\bf x}$-dependent gauge function does not affect the gauge condition
(\ref{za1}). Complete gauge fixing can be achieved for instance by imposing a
standard gauge condition on the zero modes of the neutral gluons
($\int_{0}^{\beta} dx_{0} A_{i}^{3}$) as in Ref.~\cite{LETH98}. For the
investigation of the singular fields it is advantageous not to perform this
final step in gauge fixing and to use the residual gauge freedom to change the
location of the strings. Indeed, a time-independent Abelian gauge
transformation corresponds to a redefinition of $\Omega_{{\rm D}}$,
\begin{equation}
  \label{za6}
  \Omega_{{\rm D}}\left({\bf x}\right) \rightarrow
  e^{i\tau_{3}\psi\left({\bf x}\right) }\Omega_{{\rm D}}\left({\bf x}\right) 
\end{equation}
with  arbitrary $\psi$. The related changes in the singular gauge fields are
\begin{eqnarray}
  a_{i}^{{\rm s}}\left({\bf x}\right)  & \rightarrow  & 
  a_{i}^{{\rm s}}\left({\bf x}\right) 
  - \frac{1}{g} \partial_{i}\psi\left({\bf x}\right) , \nonumber\\
  \phi_{i}^{{\rm s}}\left({\bf x}\right) & \rightarrow  &
  \phi_{i}^{{\rm s}}\left({\bf x}\right)e^{2 i\psi\left({\bf x}\right)} .
\label{am15}
\end{eqnarray}
Using this ambiguity in the diagonalization of the Polyakov loop, i.e.,
the residual gauge freedom, strings can be deformed in space. For
instance, the gauge function
\begin{equation}
\psi\left({\bf x}\right)= -\varphi  \left({\bf x}\right) 
\label{am16} 
\end{equation}
changes a string along $\theta ({\bf x})= 0 $ into a string
along $\theta ({\bf x})= \pi$. 
More generally, the gauge function which transforms a gauge field with
string ${\bf s}_{1}({\bf x})$ into a field with string
${\bf s}_{2}({\bf x})$ can be constructed by using identities
well known from magnetostatics and is given by \cite{FRHR77}
\begin{equation}
  \label{za6a}
  \psi({\bf x}) = \int _{\Sigma({\bf s}_{1}, {\bf s}_{2})}
  \frac{{\bf x}- {\bf x}^{\prime}}{|{\bf x}- {\bf x}^{\prime}|^3}
  \cdot d^2\bbox{\sigma}({\bf x}^{\prime})
  ,
\end{equation}
where the integral is performed over an arbitrary surface bounded by
the two strings (and possibly extending to infinity). 

Topological considerations \cite{KRSW87} can be used to characterize some (but
not all; cf.\ Sec.~\ref{parameters}) of the properties of the singularities
from a more general point of view.  The residual Abelian symmetry provides the
basis for this discussion.  The ambiguity in the definition of
$\Omega_{{\rm D}}$ expressed by Eq.~(\ref{za6}) suggests one can interpret
$\Omega_{{\rm D}}$ as an element of $\text{SU}(2)/\text{U}(1)$.  Poles appear
where the eigenvalues of $P$ are degenerate and $\Omega_{{\rm D}}$ is
ill-defined. Due to the non-triviality of the second homotopy group
$\pi_2(\text{SU}(2)/\text{U}(1))=Z$ a winding number can be assigned to this
singularity.  A configuration of $\Omega_{{\rm D}}$ on a sphere around
${\bf x}^{{\rm N},{\rm S}}$ with a non-vanishing winding number is not smoothly
deformable to a constant and a discontinuity will remain upon shrinking the
sphere to a point.  The winding number of $\Omega_{{\rm D}}$ around
${\bf x}^{{\rm N},{\rm S}}$ can be expressed as
\begin{eqnarray}
  \label{winding-number}
  n &\equiv& \frac{i}{4\pi} \epsilon_{ikl}
  \int_{|{\bf x}-{\bf x}^{{\rm N},{\rm S}}|=\varepsilon} d^2\sigma_i 
  {\rm tr}\, \tau_3 \, \Omega_{{\rm D}} \partial_k \Omega_{{\rm D}}^\dagger \,
  \Omega_{{\rm D}} \partial_l \Omega_{{\rm D}}^\dagger \nonumber\\
  &=& -\frac{i}{4\pi} \epsilon_{ikl} 
  \int_{|{\bf x}-{\bf x}^{{\rm N},{\rm S}}|=\varepsilon} d^2\sigma_i \,
  \partial_k (\Omega_{{\rm D}} \partial_l \Omega_{{\rm D}}^\dagger)
  .
\end{eqnarray}
As it should be, $n$ is invariant under multiplication of $\Omega_{{\rm D}}$
with an (${\bf x}$ dependent) diagonal matrix from the left. With the
winding number density being a total derivative, $n$ can be different
from zero only if $\Omega_{{\rm D}}$ is discontinuous at some point
${\bf x}^0$ on any sphere surrounding ${\bf x}^{{\rm N},{\rm S}}$.  These
points will form a string of discontinuity emanating from
${\bf x}^{{\rm N},{\rm S}}$.  The discontinuity is due to the undetermined
$\text{U}(1)$ factor of $\Omega_{{\rm D}}$.  When lifting a function from
$\text{S}^2$ to $\text{SU}(2)/\text{U}(1)$ having a non-vanishing winding-number
to $\text{SU}(2)$, points with an Abelian discontinuity can not be avoided
\cite{BLAU95}:
\begin{equation}
  \Omega_{{\rm D}}({{\bf x}}) \to e^{i\psi({{\bf x}})\tau_3} \Omega_{{\rm D}}^0
  \qquad\text{for}\quad {{\bf x}}\to {\bf x}^0 
  .
\end{equation}
Here, $\psi$ changes by an integer multiple of $2\pi$ when the string
is encircled once.  We may think of this integer as a winding number assigned to
the string,
\begin{equation}
  \label{winding-number-string}
  n^{\text{string}} \equiv \frac{\delta\psi}{2\pi}
  = \frac{1}{2\pi} \int_{C} d s_j \,
  \partial_j \psi
  = \frac{-i}{4\pi} \int_{C} d s_j \,
  {\rm tr}\, \tau_3 \Omega_{{\rm D}} \partial_j \Omega_{{\rm D}}^\dagger
  ,
\end{equation}
where the integrals are performed over a curve $C$ around the string.  By
Eq.~(\ref{winding-number}), this winding number, or if there are several
discontinuities the sum of the respective winding numbers, equals the winding
number of the monopole $n$.

For our particular parametrization, Eq.~(\ref{am3}), $\Omega_{{\rm D}}$
becomes discontinuous when approaching the curve $\theta=0$,
\begin{equation}
  \Omega_{{\rm D}}({\bf x}) \to e^{i\varphi({\bf x})\tau_3} 
  \left(
    \begin{array}{cc}
      1&0\\
      0&1\\
    \end{array}
  \right)
  \qquad\text{for}\quad \theta\to0
  .
\end{equation}

Note that the Abelian magnetic field associated with the  singular 
field $a_{i}^{{\rm s}}$ (cf.\ Eqs.\ (\ref{am5a},~\ref{am6})),
\begin{equation}
  b_k^{{\rm s}} = \epsilon_{kij} \partial_i a_j^{{\rm s}} 
  = \epsilon_{kij} \partial_i \left( \frac{1}{2 i g} {\rm tr}\, \tau_3 \,
  \Omega_{{\rm D}} \partial_j \Omega_{{\rm D}}^{\dagger} \right)
  ,
\end{equation}
is proportional to the integrand in Eq.~(\ref{winding-number}). Thus, all
topologically non-trivial point singularities give rise to magnetic monopoles
with quantized charges $2\pi n/g$.  These singularities are therefore
characterized by two charges, the magnetic charge and the north-south charge
(cf.\ Eq.~(\ref{za2})).  The emergence of two quantum numbers is a specific
property of our gauge choice which involves diagonalization of an element of
the group rather than the Lie algebra.  The presence of two quantum numbers
will be seen to be crucial for establishing a relation to the topological
charge in Sec.~\ref{instantons}.

\section{Continuous and Singular Fields in Axial Gauge}
\label{fields}

In this section we construct the singular gauge field configurations in the
axial gauge. To this end, we have to account for the other two factors
appearing in the gauge fixing transformation of Eq.~(\ref{cs2}) besides the
diagonalization ($\Omega_{{\rm D}}$) of the Polyakov loops. This construction
allows one to decompose the gauge field $A_{i}^{\prime}$ into regular and
singular terms which separately are periodic in time.  Such a decomposition is
a prerequisite for formulating the (gauge fixed) path integral in the presence
of singular field configurations. It is obvious that for such a sum of regular
and singular terms to correspond to a continuous field before gauge fixing, the
regular term has to satisfy certain conditions in the neighborhood of the
singularities which will be derived now. The starting point is the continuity
of the original field configuration. The first step in the gauge fixing
procedure does not produce any singularities,
\begin{equation}
  \label{A-Weyl}
  A_i^{(1)} \equiv P e^{ig\int_{0}^{x_{0}} dt A_{0}(t,{\bf x})}
  \left(A_i + \frac{1}{i g} \partial_i\right)
  \left(P e^{ig\int_{0}^{x_{0}} dt A_{0}(t,{\bf x})}\right)^{\dagger}
  = \text{continuous} 
  ,
\end{equation}
because the path ordered integral is continuous in ${\bf x}$ (but
not periodic in $x_{0}$).  The remaining steps can be written as
\begin{equation}
  \label{gf4}
  A_i^{\prime}
  = e^{-iga_{0}x_{0}\tau_{3}} \Omega_{{\rm D}} A_{i}^{(1)}
  \Omega_{{\rm D}}^{\dagger} e^{iga_{0}x_{0}\tau_{3}}
  + e^{-iga_{0}x_{0}\tau_{3}} s_{i} e^{iga_{0}x_{0}\tau_{3}}
  + \partial_{i} a_{0} x_{0} \tau_{3} 
  ,
\end{equation}     
where we have pulled the second factor in Eq.~(\ref{cs2}) through
$\Omega_{{\rm D}}$ and inserted the singular field $s_i$ arising from the
diagonalization (cf.~Eq.~(\ref{am5a})).  We observe that the
off-diagonal part of the singular field acquires a time dependent (and
non-periodic) phase $2 g a_0 x_0$.  At the north pole, this phase
vanishes and does not influence the divergent part of $s_i$.  At the
south pole, however, $ga_0\to\pi/\beta$, and a time-dependent (but now
periodic) phase persists.  Since we do not want to introduce a
non-periodic singular field, we decompose the gauge field as
\begin{equation}
  \label{gf13}
  A_i^{\prime} = \bar{\omega} s_i \bar{\omega}^{\dagger} + \hat A_i
  ,
\end{equation}
where $\bar{\omega}$ is the diagonal matrix 
\begin{equation} 
\bar{\omega}(x) = \left (\begin {array}{cc} \omega^{1/2}(x) & 0
\\ 0 & \omega^{-1/2}(x) \end {array}
\right )
\label{gf16a} 
\end{equation}
with $\omega(x)$ satisfying
\begin{eqnarray}
  \label{gf16}
  \omega(x) \to & 1 & \quad\text{at northern monopole} , \nonumber\\
  \omega(x) \to & e^{-2i\pi x_{0}/\beta} & \quad\text{at southern monopole} .
\end{eqnarray}
With this choice, $\hat A_i$ is finite.  To investigate its
discontinuity, we consider
\begin{eqnarray}
  \label{cont-I}
  \lefteqn{\Omega_{{\rm D}}^{\dagger} e^{iga_{0}x_{0}\tau_{3}} \hat{A}_{i}
  e^{-iga_{0}x_{0}\tau_{3}} \Omega_{{\rm D}} } \qquad \nonumber \\
  &=& A_i^{(1)} + \Omega_{{\rm D}}^{\dagger} 
  \left(s_i - e^{iga_{0}x_{0}\tau_{3}} \bar{\omega} s_i
    \bar{\omega}^{\dagger} e^{-iga_{0}x_{0}\tau_{3}}
    + \partial_{i} a_{0} x_{0} \tau_{3} \right) \Omega_{{\rm D}} \nonumber\\
  &=& A_i^{(1)} + \Omega_{{\rm D}}^{\dagger} 
  \left(\begin{array}{cc}
      \partial_{i} a_{0} x_{0} &
      \phi_i^{{\rm s}} (1 - e^{2iga_{0}x_{0}} \omega) \\
      {\phi_i^{{\rm s}}}^{\dagger}
      (1 - e^{-2iga_{0}x_{0}} \omega^{\dagger})
      & -\partial_{i} a_{0} x_{0} 
    \end{array}\right)
  \Omega_{{\rm D}}
  . 
\end{eqnarray}
The right-hand side of this equation is continuous at the strings because there
the discontinuities of $\phi^{{\rm s}}_i$ and $\partial_{i}a_0$ are canceled
by the discontinuity of $\Omega_{{\rm D}}$.
An expansion in terms of $g\beta a_0$ or $\pi-g\beta a_0$ yields that
it is also continuous at the monopoles.
On the left-hand side, we can replace $\exp(iga_0x_0\tau_3)$ by $\bar{\omega}$.
The final continuity requirement for $\hat{A}_i$ reads
\begin{equation}
  \label{cont-req}
  \Omega_{{\rm D}}^{\dagger} \bar{\omega}^{\dagger} \hat{A}_{i}
  \bar{\omega} \Omega_{{\rm D}} = \text{continuous}
  .
\end{equation}
The discontinuous factor $\Omega_{{\rm D}}$ implies a discontinuity in
$\hat{A}_i$ which cannot be absorbed into a redefinition of the inhomogeneous
singular terms $s_{i}$ in Eq.~(\ref{gf13}). For the following it is convenient
to introduce a spherical color basis,
\begin{eqnarray}
  \label{gf12a}
  \hat{A}_{i}^{1} & = &
  \frac{1}{\sqrt{2}} (\hat{\Phi}_{i}+\hat{\Phi}_{i}^{\dagger}) ,
  \nonumber\\
  \hat{A}_{i}^{2} & = &
  \frac{i}{\sqrt{2}} (\hat{\Phi}_{i}-\hat{\Phi}_{i}^{\dagger}) .
\end{eqnarray}
In this basis the continuity requirement reads
\begin{eqnarray}
  \label{gf15}
  \cos\theta\, \hat{A}_{i}^{3}
  - \sin\theta \left[e^{-i\varphi  }
      \frac{1}{\sqrt{2}} \hat{\Phi}_{i} \omega^{\dagger}
      + \text{c.c.}\right]
    &=& \text{continuous} , \nonumber \\
  \sin\theta\, e^{-i\varphi  } \hat{A}_{i}^{3}
  + (1+ \cos{\theta }) e^{-2i\varphi  } 
    \frac{1}{\sqrt{2}} \hat{\Phi}_{i} \omega^{\dagger}
  - (1- \cos{\theta }) \frac{1}{\sqrt{2}}
    \hat{\Phi}_{i}^{\dagger} \omega  &=&  
\text{continuous} ,\label{gf15a} 
\end{eqnarray}
which determines the behavior of the fields $\hat{A}_{i}$ in the neighborhood
of monopoles and strings. In particular, along the strings, Eq.~(\ref{gf15a})
reduces to
\begin{equation}
  \label{gf17}
 e^{-2i\varphi  } \hat{\Phi}_{i} = \text{continuous at the string},  
\end{equation}
a condition familiar from the treatment of the motion of a charged particle in
the field of a Dirac monopole. Continuity of the original field configuration
guarantees ``invisibility'' of strings and monopoles. These continuity
requirements clearly differentiate between charged and neutral components. In
particular, the neutral fields $\hat{A}_{i}^{3}$ are not affected by the
presence of the strings. Abelian dominance, though not valid for the singular
configurations, possibly has its origin in this distinction and thus would be a
property of the fluctuating fields.  \newline On the basis of the decomposition
(\ref{gf13}), the field strength associated with singular field configurations
is straightforwardly calculated.  The definition of charged gluon fields in the
spherical color basis of Eq.~(\ref{gf12a}) suggests introduction, alongside the
neutral component
\begin{equation}
  \label{gf18}
  F_{ij}^{3} = \partial_{i}A_{j}^{3}- \partial_{j}A_{i}^{3}-
ig(\Phi_{i}^{\dagger}\Phi_{j}-\Phi_{j}^{\dagger}\Phi_{i}) ,
\end{equation}
of the following charged component of the field strength,
\begin{equation}
  \label{gf19}
  \chi_{ij} = (\partial_{i} + ig A_{i}^{3})\Phi_{j}-
  (\partial_{j} + ig A_{j}^{3})\Phi_{i} .
\end{equation}
The spatial part of the Yang-Mills Lagrangian is then given by
\begin{equation}
  \label{gf20}
  \frac{1}{2}\sum_{a=1}^{3} F_{ij}^{a} F_{ij}^{a} =  \frac{1}{2}
  F^{3}_{ij} F_{ij}^{3}+\chi_{ij} \chi_{ij}^{\dagger} . 
\end{equation}

It is furthermore convenient to include the Polyakov loop variables 
$a_{0}({\bf x})$ into the definition of the singular field configuration. In
this way, the dependence on the original variable $A_{0}$ is completely
absorbed into the singular part $\alpha_{\mu}$ defined by
\begin{eqnarray}
  \label{gf22}
  \alpha_{\mu}^{3}({\bf x})
  & = & 2\left(a_{\mu}^{{\rm s}}({\bf x})(1-\delta_{\mu,0}) 
    + a_{0}({\bf x})\delta_{\mu,0}\right) , \nonumber \\
  \alpha_{\mu}^{1}(x)+i\alpha_{\mu}^{2}(x) 
  & = & 2 \omega(x) \phi_{\mu}^{{\rm s}}({\bf x}) (1-\delta_{\mu,0}) .
 \end{eqnarray}
The total field is given by 
\begin{equation}
  \label{gf23}
  A^{\prime}_{\mu} = \alpha_{\mu} + \hat{A}_{\mu},
\end{equation}
where in axial gauge $\hat{A}_{0}=0$. The field strength is accordingly
decomposed,
\begin{equation}
  F_{\mu\nu}[\hat{A}+\alpha] 
  = F_{\mu\nu}[ \alpha ] 
  + \hat{D}_{\mu}\hat{A}_{\nu} - \hat{D}_{\nu}\hat{A}_{\mu} 
  + ig[\hat{A}_{\mu},\hat{A}_{\nu}]
\label{gf2422} 
\end{equation}
with the covariant derivative
\begin{equation}
\hat{D}_{\mu} = \partial_{\mu} +ig[\alpha_{\mu}, 
\label{gf25} 
\end{equation}
being defined in terms of the singular fields. For the evaluation of
the field strength we observe that for $\omega(x)\equiv 1$
(cf.~Eq.~(\ref{gf16})) the singular gauge field $\alpha_{i}$ is a
spatial pure gauge, i.e., the spatial field strength components vanish
in this limit. This simplifies the expressions and we easily find
\begin{eqnarray}
  \label{gf26}
  F_{i 0}^{3}\left[\alpha\right] 
  & = & 2\partial_{i}a_{0}({\bf x}) , \qquad  
  \chi_{i 0}[s] = \sqrt{2} \phi_{i}^{{\rm s}}({\bf x})
  (-\partial_{0}+2iga_{0}({\bf x})) \omega(x) , \nonumber\\  
  F_{ij}^{3}\left[\alpha\right] 
  & = & 2ig(\omega(x)\omega(x)^{\dagger}-1)
  (\phi_{i}^{{\rm s}\,\dagger}({\bf x})\phi_{j}^{{\rm s}}({\bf x})
  - \phi_{j}^{{\rm s}\,\dagger}({\bf x})\phi_{i}^{{\rm s}}({\bf x})) ,
  \nonumber\\
  \chi_{ij}[\alpha] & = & 
  \sqrt{2} (\phi_{j}^{{\rm s}}({\bf x})\partial_{i}\omega(x)- 
  \phi_{i}^{{\rm s}}({\bf x})\partial_{j}\omega(x)) . 
\end{eqnarray}
As could be expected, these explicit formulae for the Yang-Mills field strength
components display that in general no particularly large or even divergent
action is associated with singular gauge field configurations, i.e., monopoles
generated by gauge fixing have finite action.  More subtle in these expressions
for the field strength is the distinction between configurations which contain
one type of singularity (north or south poles only) from configurations in
which both types occur as a function of ${\bf x}$.  In the former case we can
choose $\omega(x)= 1$ for configurations with north and $\omega(x)= \exp(2i\pi
x_{0}/\beta)$ for configurations with south pole singularities only and the
action generated by such configurations can be arbitrarily small. In this case
the spatial components of the associated field strength vanish, $F_{ij} = 0$.
If both types of singularity are present, there are necessarily spatial
components which contribute to the action.  As we shall see below, instantons
belong to field configurations of this second type.  Fluctuations around the
singular fields are described by the fields $\hat{A}_{\mu}$ in
Eq.~(\ref{gf23}).  The covariant derivative $\hat{D}_{\mu}$ of Eq.~(\ref{gf25})
couples these two fields. The boundary conditions (\ref{gf15a}) imposed on the
fluctuating fields guarantee finiteness of the total action. The relevance of
these expressions for formulation of the quantum theory in the presence of
singular field configurations is obvious.

Finally, we point out the residual local gauge invariance of the action 
\begin{equation}
S[\hat{A}+\alpha] = \frac{1}{2}\int d^{4}x  \,{\rm tr} 
\left(F_{\mu\nu}[\hat{A}+\alpha ] F_{\mu\nu}[\hat{A}+\alpha
  ]
 \right) .  
\label{am24a} 
\end{equation}
The effect on $S$ of rotating  the color components,
\begin{equation}
  \hat{A}_{\mu}^{\prime} \left(x\right) = e^{i\psi \left({\bf x}\right)\tau_{3}}
  \hat{A}_{\mu}  \left(x\right) e^{-i\psi \left({\bf x}\right)\tau_{3}} ,
\label{am24b} 
\end{equation}
can be compensated for by an appropriate change in the singular background field
(cf.~Eq.~(\ref{am15})),
\begin{equation}
  \alpha^{\prime}_{\mu} = e^{i\psi \left({\bf x}\right)\tau_{3}}\left(
    \alpha_{\mu} +\frac{1}{ig}\partial_{\mu}\right)
  e^{-i\psi \left({\bf x}\right)\tau_{3}} .
\label{am24c} 
\end{equation}

\section{Monopole Parameters}
\label{parameters}

In the completely gauge fixed formulation there can be no dependence on the
orientation of the Polyakov loops (or, equivalently, of $\tilde{\bf{a}}$).  This
is not manifest in our formulation since the expressions for the singular field
configurations (\ref{gf14}) depend explicitly on the orientation of
$\tilde{\bf{a}}$.  Quantities like the Abelian magnetic field which are
invariant under residual gauge transformations (Eqs.~(\ref{am15},~\ref{am24c}))
are, however, independent of these redundant orientational variables.  For the
following discussion, we focus on fields in the neighborhood of a monopole
which we assume to be located at the origin. Once more, the starting point is
the regularity of the Polyakov-loop variables $\tilde{\bf{a}}$ before gauge
fixing which allows for the Taylor expansion
 \begin{equation}
  \label{mp1}
  \tilde{\bf{a}} = M  {\bf x} + \ldots \;.
\end{equation}
In polar decomposition, the matrix $M$ is written as 
\begin{equation}
  \label{mp2}
  M= \rho m
\end{equation}
with a symmetric matrix $m$ and an orthogonal matrix $\rho$ with
positive determinant.  We have
\begin{equation}
  \label{mp3}
  \frac{\partial^{2}\sin^{2}(g\beta a_{0})}{\partial x_{i}\partial x_{j}}
  = 2(M^{T}M)_{ij} = 2 (m^{2})_{ij}.  
\end{equation}
Thus, the matrix $m$ is given in terms of the gauge invariant variables
$a_{0}({\bf x})$.  Close to the origin, the Abelian magnetic field of the
monopole (cf.~Eq.~(\ref{gf14})) is given by
\begin{equation}
  \label{mp6}
  b_{k} = \frac{1}{2g} \epsilon_{kij} \sin \theta  \, 
  \partial_{i}\theta  \, \partial_{j}\varphi 
  \approx \frac{1}{2g}\det{m} \frac{x_{k}}{\sin^{3}(g\beta a_{0})} .
\end{equation}
It is thus determined exclusively by the properties of $a_{0}({\bf x})$. 

More generally, the matrix $\rho$ of the decomposition (\ref{mp2}) describes a
rotation of the color coordinate system which can be accounted for by a
residual gauge transformation (cf.~Eqs.~(\ref{am24b},~\ref{am24c})) and
therefore does not affect gauge invariant quantities like the action
(Eq.~(\ref{am24a})).  This can be seen in the following way.  For fixed $m$,
the orthogonal matrix $\rho$ can also be interpreted as a rotation of the
spatial coordinate ${\bf y}=m{\bf x}$.  If expressed in terms of ${\bf y}$, the
fields ${\bf a}^{{\rm s}}$ and $\bbox{\phi}^{{\rm s}}$ are just rotated by
$\rho$.  As mentioned above this rotation can be undone by performing an
appropriate residual gauge transformation (\ref{am15}).

Thus, in the neighborhood of the points ${\bf x}^{{\rm N},{\rm S}}$ the
singularities can be characterized completely by the gauge invariant Polyakov
loop variables $a_{0}({\bf x})$. This is trivially correct for those properties
which are determined by general topological requirements.  However our
discussion also indicates that topological arguments are not sufficient to
fully characterize the singularities. We illustrate this by rewriting the
expression (\ref{mp6}) for the magnetic field.  Integration of Eq.~(\ref{mp6})
over a closed surface $\Sigma$ around ${\bf x} = {\bf 0}$ yields the magnetic
charge
\begin{equation}
  \label{mp16}
  \int_{\Sigma} d^{2}\sigma_{i} \, b_{i} =
  {\rm sgn}(\det m) \frac{2\pi}{g} ,
\end{equation}
which, as already follows from homotopy arguments, is independent of the
details of the matrix $m$.  Note that $b_{i}$ only carries one quantum
of magnetic charge.  Higher charges can occur when $M=0$ and higher terms in
the Taylor expansion govern the behavior of $\tilde{\bf{a}}$ near the
origin.  We decompose the magnetic field as
\begin{equation}
  \label{mp17}
  b_{k} = \frac{{\rm sgn}(\det m)}{2g}
  \frac{x_{k}}{\left(\sum_{i}x_{i}^{2}\right)^{3/2}}
  + b_{k}^{\text{tr}} ,
\end{equation}
where 
\begin{equation}
  \label{mp17a}
  b_{k}^{\text{tr}}
  = \frac{x_{k}}{2g} \left[\frac{\det m}
    {\left(\sum_{ij}(m^2)_{ij} x_{i} x_{j}\right)^{3/2}}
    - \frac{{\rm sgn}(\det m)}{\left(\sum_{i}x_{i}^{2}\right)^{3/2}}
  \right]. 
\end{equation}
By construction, the magnetic charge of $b_{k}^{\text{tr}}$ vanishes,
\begin{equation}
  \label{mp18}
  \int_{\Sigma} d^{2}\sigma_{i} b_{i}^{\text{tr}}= 0 .
\end{equation}
Thus, $b_k^{\text{tr}}$ has identically vanishing divergence, i.e., it is a
transverse singular field. This transverse field is in turn determined by
currents which are generated by the charged components of the singular field
configuration (\ref{gf14}). For $\omega(x)\equiv 1$ or
$\omega(x)\equiv\exp(2i\pi x_0/\beta)$, the non-Abelian field strength has no
spatial components which implies the transverse Abelian magnetic field to
satisfy the Maxwell equation
\begin{equation}
  \label{mp18a}
  {\rm curl} \,{\bf b}^{\text{tr}}= g\,{\bf j} \quad \text{with} \quad {\bf j}
 = {\rm curl} ({\bf \Phi}^{\dagger} \times {\bf \Phi} ) .
\end{equation}

\section{Monopoles and Instantons}
\label{instantons}

Up to this point our discussion has focused on the properties of the
singularities which are generated in axial gauge if the Polyakov loop passes
through the center of the group. In this section we establish a physically
important sufficient condition for the presence of such singularities by
relating the monopole charges associated with a certain field configuration
to its topological charge. That such a relation must exist in the axial gauge
is easily understood.  As is well known, the topological charge
\begin{equation}
  \label{Pontrjagin-index}
  \nu = \frac{g^2}{16\pi^2} \int d^4x \, {\rm tr}\, F_{\mu\nu}
  \tilde F_{\mu\nu} 
  ,\quad\text{where}\quad
  \tilde F_{\mu\nu} = \frac12 \epsilon_{\mu\nu\rho\sigma} F_{\rho\sigma} 
  ,
\end{equation}
can be written as a surface integral,
\begin{equation}\label{nu-surface}
  \nu = \frac{g^2}{4\pi^2} 
  \int d^3\Sigma_{\mu}\, K_{\mu} 
\end{equation}
with the topological current given by
\begin{equation}\label{K}
  K_{\mu} = \epsilon_{\mu\nu\rho\sigma} \, {\rm tr} \left(
    \frac14 A_{\nu} F_{\rho\sigma}
    + \frac{i g}{6} A_\nu A_\rho A_\sigma \right) .
\end{equation}
In the axial gauge representation, the fields satisfy Abelian boundary
conditions (see below) and a finite topological charge can thus arise
only if singularities are present. In this case $\nu$ receives its
full contribution from a surface surrounding the singularities.

More technically, to establish this relation we assume space to be a
three-dimensional sphere.  Generalizations to other compact manifolds or
$\text{R}^3$ are possible.  We may cover $\text{S}^3$ by two charts whose
overlap does not contain any singularities (and where $P\ne\pm1$).  In order
for the gauge fields on both charts to fulfill the gauge condition (\ref{za1}),
the transition function relating the two charts must be Abelian and
time independent and can therefore be continuously deformed to unity.  The
interpolating function can be used to transform the transition function away.
Since the fields are furthermore periodic in time in the gauge chosen, we can,
apart from the singularities, assume trivial boundary conditions.  We
furthermore assume that only generic singularities occur, i.e., points ${\bf
  x}_i$ where $P({\bf x}_i)=z_i=\pm 1$ and strings ${\bf x}_{\alpha}(s)$ with
$0\le s\le 1$ which may either connect two singular points ${\bf x}_{i}$ or be
closed.  In the volume formula for the Pontrjagin index
(\ref{Pontrjagin-index}) we then exclude small regions around the
singularities.  The correct result will be recovered after shrinking these
regions to zero volume, because the non-Abelian field strength is finite
everywhere.  In the surface integral (\ref{nu-surface}), only the boundaries of
the excluded regions will appear.

Because the singularities are time independent, $\mu$ in Eq.~(\ref{nu-surface})
is a spatial index.  The second term in Eq.~(\ref{K}) therefore necessarily
involves $a_0$,
\begin{equation}\label{KAAA}
  K^{A A A}_j
  = -\frac{i g}{2} \epsilon_{jkl} \, a_{0} \,{\rm tr}\, \tau_3 A_k A_l .
\end{equation}
Under the trace, only the off-diagonal components of $A_k$ and $A_l$
contribute.  These are finite at the strings and diverge linearly at
the monopoles (cf.~Eqs.~(\ref{gf13},~\ref{am6},~\ref{gf14})).  At the
monopoles only the quadratic singularity generated by the divergent
parts of both field components contributes,
\begin{eqnarray}
  \nu^{A A A} &=& \frac{g^2}{4\pi^2} \int d^3\sigma_j\, K^{A A A}_j
  \nonumber\\
  &=& -\frac{i g}{8 \pi^2} \epsilon_{jkl} \int d^3\Sigma_j \, a_{0}
  \,{\rm tr}\, \tau_3 \, 
  \bar{\omega} \Omega_{{\rm D}} \partial_k \Omega_{{\rm D}}^\dagger
  \bar{\omega}^\dagger \,
  \bar{\omega} \Omega_{{\rm D}} \partial_l \Omega_{{\rm D}}^\dagger
  \bar{\omega}^\dagger .
\end{eqnarray}
The diagonal matrices $\bar\omega$ and $\bar\omega^{\dagger}$ cancel
under the trace.  The integration over surfaces oriented towards the
monopoles splits into a (trivial) time integral and an integral over
spheres around monopoles, which we choose to be oriented outwards:
\begin{eqnarray}
  \nu^{A A A} 
  &=& -\sum_i \frac{i g \beta a_{0}({\bf x}_i)}{8\pi^2} \epsilon_{jkl}
  \int_{|{\bf x}-{\bf x}_i|=\varepsilon}
  d^2\sigma_j \, {\rm tr}\, \tau_3 \,
  \Omega_{{\rm D}}\partial_k \Omega_{{\rm D}}^\dagger \,
  \Omega_{{\rm D}} \partial_l \Omega_{{\rm D}}^\dagger \nonumber\\
  &=& -\frac12 \sum_{\scriptstyle i\atop\scriptstyle P({\bf x}_i)=-1} n_i
  ,
  \label{nu-AAA}
\end{eqnarray}
where we have identified the monopole winding number (\ref{winding-number}).

In the first term of Eq.~(\ref{K}), the field strength is finite
because it transforms homogeneously under gauge transformations.
Therefore, this term can at most diverge linearly and will contribute
only at the strings.  There, only the diagonal part of $A_{\nu}$
is singular (cf.~Eq.~(\ref{gf14})).  In the field strength $F_{0
  l}=\partial_{0} A_l-\partial_l a_{0} \tau_3 -i g[a_{0} \tau_3,A_l]$,
the commutator is purely off-diagonal and the first term integrates to
zero because $A_l$ is periodic and the singular fields are time
independent.  Therefore, the contribution to the winding number reads
\begin{equation}
  \nu^{A F} = - \frac{g}{8 i \pi^2} \epsilon_{j k l} \int d^3\Sigma_j \,
  {\rm tr} \, \tau_3 \, 
  \bar{\omega} \Omega_{{\rm D}} \partial_k \Omega_{{\rm D}}^\dagger
  \bar{\omega}^{\dagger} \, \partial_l a_{0} .
\end{equation}
Again, $\bar{\omega}$ and $\bar{\omega}^{\dagger}$ drop out.  Inserting the
explicit expression for the diagonal singular field, Eq.~(\ref{gf14}), we
obtain
\begin{eqnarray}
  \nu^{A F} 
  &=& - \frac{g}{8\pi^2} \epsilon_{j k l} \int d^3\Sigma_j\,
  (1+\cos\theta) \, \partial_k \varphi \, \partial_l a_{0}
  \nonumber\\
  &=& \frac{g \beta}{2\pi} \sum_\alpha n_\alpha
  \Bigl(a_{0}({\bf x}_{\alpha}(1))-a_{0}({\bf x}_{\alpha}(0))\Bigr) 
  ,
  \label{nu-AF}
\end{eqnarray}
where the winding number carried by the strings,
Eq.~(\ref{winding-number-string}), has been inserted.  Equation (\ref{nu-AF})
shows that closed strings, for which ${\bf x}_{\alpha}(1)={\bf x}_{\alpha}(0)$,
do not contribute to the Pontrjagin index. Open strings must emanate from (end
on) monopoles of the same (opposite) charge as the string. Therefore,
Eq.~(\ref{nu-AF}) gives the same contribution as Eq.~(\ref{nu-AAA}).  If we
further make use of the fact that on a compact manifold the total magnetic
charge must vanish,\footnote{Non-trivial boundary conditions have been
  converted to singularities.} we can split this into contributions from
northern and southern poles,
\begin{equation}
  \label{nu-final}
  \nu = \frac12 \Biggl( 
    \sum_{\scriptstyle i\atop\scriptstyle P({\bf x}_i)=1} n_i 
    - \sum_{\scriptstyle i\atop\scriptstyle P({\bf x}_i)=-1} n_i 
  \Biggr) 
  = \frac12 \sum_i z_i n_i ,
\end{equation}
i.e., the Pontrjagin index is determined by the two charges characterizing the
monopoles (cf.\ Eqs.~(\ref{za2}) and (\ref{winding-number})) and 
is given by the difference
in the total charge of north ($P=1$) and of south poles ($P=-1$).
Thus, an instanton configuration or a continuous deformation thereof
must contain at least two magnetic monopoles.  Indeed, it is
straightforward to verify this relation for a periodic
(finite-temperature) instanton.  Explicit expressions are given in the
next section.  On the other hand, the presence of magnetic monopoles
does not necessarily give rise to a topological charge. For instance,
configurations describing oscillations of the Polyakov loop around the
north pole which are most likely relevant for the deconfined phase
generate a succession of $P=1$ poles and antipoles and thus do not
contribute to the total charge at all. It is plausible that such
oscillations around $P=1$ can be continuously deformed to a constant
and the presence of two types of singularity is therefore required for
a non-vanishing topological charge. The relation (\ref{nu-final})
actually displays the topological origin of the difference between
field configurations containing only one or both types of singularity
which has already been observed in our discussion of the expressions
(\ref{gf26}) for the field strength.

The above relation between instanton number and axial gauge monopoles has been
derived before in a different framework \cite{Reinhardt97} in which, however,
not only the gauge field but also the non-Abelian field strength becomes
singular.  The coincidence of both results is remarkable since in our approach
it is essential that the non-Abelian field strength transforms covariantly
under the gauge fixing transformation (\ref{cs2}).

\section{Physics of Axial Gauge Monopoles -- Conclusions}
\label{physics}

We have presented a detailed analysis of the singular field configurations
which occur when representing QCD in (modified) axial (temporal) gauge. (Most
of our results actually also apply to the only partially fixed Polyakov gauge.)
Singular gauge fields emerge as coordinate singularities when diagonalizing the
(untraced) Polyakov loop or, equivalently, when defining the (local) color
coordinate system by the direction of the Polyakov loop. This coordinate choice
becomes ill-defined whenever the Polyakov loop is in the center of the group.
At these points, the gauge fixed fields necessarily develop singularities. As
is well known, homotopy arguments guarantee that, for a generic field
configuration, the neutral Abelian field strength associated with the gauge
fixed field configuration develops a monopole singularity. The complete and
gauge covariant non-Abelian field strength must remain regular in the course of
gauge fixing.  Therefore, the charged components of the gauge fixed field
configurations necessarily also develop singularities such that the
singularities in the Abelian field strength are exactly canceled by the
non-Abelian commutator. Furthermore, not only do the field configurations
exhibit singular charged in addition to the singular Abelian components, these
Abelian components in general also possess singularities in the transverse
components in addition to the longitudinal ones. The presence of these singular
transverse Abelian fields is not dictated by any topological requirement and,
accordingly, their strength is arbitrary. Their sources are currents
generated by the singular charged fields.

Although we have performed this detailed study of the structure of the singular
fields for a particular gauge choice, the arguments directly apply whenever
gauge fixing is achieved by diagonalization of a local quantity in the adjoint
representation. Locally the structure of the singular fields as derived above
remains valid for other gauges of this type.  Globally, however, the fields are
in general not independent of one of the space-time coordinates. Despite this
common structure of the singularities, the physics consequences must be
expected to be quite different for different gauge choices. This is indicated
by the connection which we have derived between monopole charges and the
instanton number. The presence of trivial boundary conditions in the gauge
fixed formulation is essential in establishing a relation between instantons
and monopoles. The necessity of the presence of two distinct types of
singularity (north and south pole) points to properties of this relation
which are specific for axial type gauges in which a group element is actually
diagonalized.  Finally, the simultaneous role of the Polyakov loop in
characterizing confined and deconfined phases on the one hand and in
determining positions and strengths of the singular gauge fields on the other,
is obviously a very particular property of the axial gauge which we shall
exploit in this concluding paragraph.

The central result of our investigations is summarized in the expression
for the axial gauge QCD partition function 
\begin{equation}
Z= \sum_{{\bf n}}Z_{{\bf n}} =
\sum_{{\bf n}}\int
D[a_{0}^{\bf n}]\int
\prod_{i=1}^{3}D[\hat{A}_{i}]
e^{-S[\hat{A}+\alpha]} .
\label{pm0a} 
\end{equation}  
The sum over field configurations has been decomposed according to the number
${\bf n}=(n_{{\rm N}},n_{{\rm S}})$ of north ($n_{{\rm N}}$) and south
($n_{{\rm S}}$) pole singularities; i.e., the path integral in $Z_{{\bf n}}$ is
performed over field configurations in which the Polyakov loop passes
$n_{{\rm N},{\rm S}}$ times through the north and south pole, respectively. For
this decomposition to be meaningful, regularization of the generating functional
is required. According to our analysis above, the singular component $\alpha$
(cf.~Eq.~(\ref{gf22})) is completely determined by the Polyakov loop variables,
\begin{equation}
  \label{pm0b}
  \alpha = \alpha \left[a_{0}^{{\bf n}}\right] .
\end{equation}
The quantum fluctuations $\hat{A}$ around the singular field configurations
give rise to infinite values of the action unless they satisfy the boundary
conditions (\ref{gf15}). In the generating functional $A_{0}$ has been
eliminated up to the gauge invariant Polyakov loop variables. Apart from
singularities, this elimination also generates a non-trivial but explicitly
calculable measure $D\left[a_{0}^{\bf n}\right]$, the Haar measure of
$\text{SU}(2)$ (cf.~\cite{LENT94}). The action is easily calculated in terms of
singular and regular fields by using the explicit expressions
(\ref{gf2422},~\ref{gf26}) for the field strength components.  (The inclusion
of quarks into the formalism is straightforward since it does not affect the
gauge fixing procedure.)

Starting from the above generating functional, a dynamical picture can be
developed by integrating out the Polyakov loop variables. Application of
techniques developed for the ${\bf n}= (0,0)$ sector in Ref.~\cite{LETH98} will
yield an effective theory of QCD with $A_{0}$ completely removed and with the
other degrees of freedom being coupled to the singular field configurations
discussed above. In a qualitative discussion we now attempt to anticipate some
of the main properties of such a formulation of QCD. We first discuss issues
concerning the number of singularities typically involved in a generic field
configuration starting with a brief discussion of the Polyakov loop associated
with instantons. This will lead to an estimate for the monopole density
associated with a dilute gas of instantons. Explicit expressions for finite
temperature instantons of size $\lambda$ are given in Ref.~\cite{HASH78},
\begin{equation}
  \label{pm7}
  A_{\mu} = \frac{1}{g}\bar{\sigma}_{\mu\nu}\partial_{\nu}
  \ln\left\{1+ \frac{(\gamma/u) \sinh u}{\cosh u-\cos v}\right\} ,
\end{equation}
where 
\begin{equation}
  \label{pm8}
  u= 2\pi |{\bf x}-{\bf x}_{0}|/\beta ,\qquad
  v= 2\pi t/\beta ,\qquad \gamma= 2(\pi \lambda/\beta)^{2}. 
\end{equation}
Using these expressions, the associated Polyakov loops are easily calculated,
\begin{equation}
  \label{pm9}
  P({\bf x}) = \exp\left\{i \pi
  \frac{\left({\bf x}-{\bf x}_{0}\right)\cdot{\bbox{\tau}}}
  {|{\bf x}-{\bf x}_{0}|} \chi(u)\right\}.
\end{equation}
The function 
\begin{equation}
  \label{pm10}
  \chi(u) = 1-\frac{(1-\gamma/u^{2})\sinh u+(\gamma/u) \cosh u}
  {\sqrt{(\cosh u+(\gamma/u) \sinh u)^{2}-1}}
\end{equation}
decreases monotonically from the value 1 at $u=0$ with increasing $u$
and vanishes at $u=\infty$.  Thus, the Polyakov loop associated with
an instanton passes through the south pole at $u=0$ (the center
of the instanton) and approaches the north pole at asymptotic
distances. 
As is well known, instantons carry one unit of topological charge which
determines the value $8\pi^{2}/g^{2}$ of the action. We have seen above that
with both north and south poles occurring in a certain field configuration the
corresponding magnetic field strength is necessarily different from zero
($|\omega(x)| \neq 1$ in Eq.~(\ref{gf26})). The necessarily finite value of the
action is minimized if the choice of $\omega(x)$ corresponds to an instanton.

Results of the instanton liquid model \cite{SCSH96} and of lattice QCD
\cite{CGHN94} suggest the presence of a finite density $n_{{\rm I}}$ of
instantons in the confined phase. In order to relate instanton and monopole
densities, we treat the instantons as independent; in this case the Polyakov
loop is given by a product over the Polyakov loops associated with single
instantons (or anti-instantons). For small instanton sizes $\lambda \ll \beta
$, the number of instantons plus anti-instantons integrated over in the time
integral of the Polyakov loop for fixed ${\bf x}$ is of the order of $\beta
n_{{\rm I}}\lambda^{3}$. On the average, the number of instantons and
anti-instantons will be the same and thus the phase accumulated in the time
integral will be given by the fluctuations and is thus expected to be of the
order of $\pi \sqrt{\beta n_{{\rm I}}\lambda^{3}}$. When changing the position
${\bf x}$ by an instanton size of $\lambda$ we expect, in the absence of
correlations between the instantons, a different value of the phase which
however is of the same order of magnitude.  Thus, in this change of ${\bf x}$
the Polyakov loop will typically pass $\sqrt{\beta n_{{\rm I}}\lambda^{3}}$
times through the center of the group which implies the following estimate for
the monopole density (counting both poles and antipoles) at low temperatures
\begin{equation}
  \label{pm11}
  n_{{\rm M}} \propto \left(\beta  n_{{\rm I}}\lambda\right)^{3/2}
  , \qquad \lambda \ll \beta ,
\end{equation}
while for high temperatures, the integral involves at most one instanton and we
therefore expect
\begin{equation}
  \label{pm12}
  n_{{\rm M}} \propto \beta  n_{{\rm I}} , \qquad \lambda \ge \beta . 
\end{equation}
Surprisingly, the result (\ref{pm11}) implies an infinite monopole density in
the zero temperature limit. Although derived on the basis of a finite value of
the instanton density, this result is to a large extent independent of the
particular model.  Let us consider more generally a field configuration with
significant variations on length scales $\ge \lambda_{1} $ before gauge fixing.
Using similar arguments as above, such a configuration will typically lead to a
value of $\pi\sqrt{\beta/\lambda_{1}}$ for the phase of the Polyakov loop and
if the field configurations are correlated over distances $\lambda_{2}$ the
value $(\beta/\lambda_{1}\lambda_{2}^{2})^{3/2}$ for the monopole density
follows.  The characteristic $\beta^{3/2}$ dependence is mostly a consequence
of the integration over the time interval which becomes ill-defined in the zero
temperature limit. It thus appears that irrespective of the temperature, axial
gauge monopoles condense. Even a gas of smooth field configurations with small
values of the action gives rise to a finite and in the zero temperature limit
divergent monopole density.
 
Beyond condensation of magnetic monopoles via instantons, the system has the
additional option of condensation involving one type (north or south) of poles
and corresponding antipoles only. As our discussion of the singularity
structure shows, such field configurations do not require non-vanishing
magnetic field strength and the electric field contribution to the action can
be made arbitrarily small for sufficiently smooth Polyakov loop variables. Thus,
suppression of such field configurations could only come through their coupling
to the quantum fluctuations. Since it appears that instantons may not be able
to account for confinement \cite{CADG78} or, more precisely, for a realistic
value of the string constant \cite{CGHN94}, this additional option could be
relevant and, by a correspondingly increased monopole density, generate the
missing strength for the dual Meissner effect.
 
In axial gauge, condensation of monopoles must be expected to persist beyond
the deconfinement transition. With the center symmetry broken, the Polyakov
loop is not distributed symmetrically around the equator of $\text{S}^{3}$ any
more.  It rather approaches more and more either the north or the south pole
with increasing temperature.  An expectation value $P\neq \pm 1$ in
the infinite temperature limit is compatible neither with the Stefan-Boltzmann
law (cf.~\cite{EKLT97}) nor with the expected dimensional reduction to 
2+1-dimensional QCD (cf.~\cite{LETH98}). Thus, as the Polyakov loop approaches 
one of the poles, the probability to pass through this pole and therefore the
monopole density must be expected to increase. On the other hand, for this
increased density to be compatible with perturbation theory and, in particular,
not to lead to confinement, one might expect poles and antipoles to compensate
each other to a large extent. This would be the case if poles and antipoles are
strongly correlated with each other. We thus expect the high-temperature phase
to consist of a gas of magnetic dipoles.
   
So far our qualitative arguments have dealt mainly with the entropy associated
with a finite monopole density. Dynamics has been involved only implicitly when
invoking the instanton gas picture. In concluding we now present qualitative
arguments concerning basic dynamical issues. As our formal development shows,
singular monopole configurations can have vanishing magnetic field energy and
may contribute only little to the action (cf.~Eq.~(\ref{gf26})). For this
reason we focus on the coupling of the quantum fluctuations to the singular
fields which is contained in the covariant derivative in Eq.~(\ref{gf25}). In
the presence of singularities quantum fluctuations have to satisfy the
conditions (\ref{gf15a}), otherwise infinite action results; this condition
just expresses finiteness of the action before gauge fixing. Given a finite
monopole density, and assuming no particular correlations between the
monopoles, this condition cannot be satisfied for long-wavelength
fluctuations, i.e., we expect fluctuations with wave number
\begin{equation}
  \label{pm13}
  k \le k_{\text{min}} = n_{{\rm M}}^{1/3} 
\end{equation}
to be dynamically suppressed. On the other hand, long-wavelength excitations
associated with bilinears such as ${\bf \Phi}^{\dagger}(x)\cdot{\bf \Phi}(x)$
which might be interpreted as two-gluon states are not necessarily affected by
the continuity conditions and may be associated with excitation energies which
remain finite at $T=0$. Thus, a high density of monopoles indeed seems to
generate confinementlike phenomena. At high temperatures, appropriate
correlations between poles and antipoles must make this mechanism ineffective.

In order to further illustrate effects of the coupling to magnetic monopoles,
we consider the following contribution to the action generated by the coupling
of the charged quantum fluctuations to the Abelian monopole fields via the
four-gluon vertex,
\begin{equation}
  \label{pm1}
  \delta S = -g^{2}\int d^{4}x \sum_{i=1}^{3}
  \hat{{\bf \Phi}}^{\dagger}(x) \cdot \hat{{\bf \Phi}}(x) \,
  {\bf a}^{{\rm s}}({\bf x})\cdot {\bf a}^{{\rm s}}({\bf x}) .
\end{equation}
At long wavelengths the quantum fluctuations thus acquire a mass $\delta m$ by
coupling to the Abelian magnetic field,
\begin{equation}
  \label{pm2}
  \delta m^{2} = g^{2}\frac{1}{V}\int_{V} d^{3}x \,
  {\bf a}^{{\rm s}}({\bf x}) \cdot {\bf a}^{{\rm s}}({\bf x}) .     
\end{equation}
To compute this mass term we assume the singular field
${\bf a}^{{\rm s}}({\bf x})$ to be given by a superposition of the standard
monopole fields of Eq.~(\ref{am18}),
\begin{equation}
  {\bf a}^{{\rm s}}({\bf x})
  = \sum_{i=1}^{N} {\bf a}^{n_{i}} \left({\bf x}-{\bf x}_{i}\right) 
  \label{pm3}
\end{equation}
with positions ${\bf x}_{i}$ of the monopoles and  charges $n_{i} = \pm 1$ and
vanishing total magnetic charge,
\begin{equation}
  \label{pm4}
  \sum_{i=1}^{N} n_{i} = 0 .
\end{equation}
In this way the functional integral over the Polyakov loop variables in
Eq.~(\ref{pm0a}) has effectively been replaced by a summation over positions 
${\bf x}_{i}$ and charges $n_{i}$ of these prescribed representative fields.
 
Using standard identities from electrostatics, the mass term can be written as
\begin{equation}
  \label{pm5}
  \delta m^{2} = \frac{1}{4 V}\int_{V} d^{3}x
  \sum_{i,j=1}^{N}\frac{n_{i}n_{j}}{|{\bf x}-{\bf x}_{i}||{\bf x}-{\bf x}_{j}|}
  = -\frac{\pi}{ V}\sum_{{\scriptstyle i,j=1\atop\scriptstyle i<j}}^{N}
  n_{i}n_{j}|{\bf x}_{i}-{\bf x}_{j}| .     
\end{equation}

Similarly, neutral gauge field fluctuations acquire a mass $\delta m_3$ by
coupling to the singular charged components,
\begin{equation}
  \label{pm6}
  \delta m^{2}_{3} = g^{2}\frac{1}{V}\int_{V} d^{3}x \,
  {\bbox{\phi}}^{{\rm s}\,\dagger} ({\bf x}) \cdot
  {\bbox{\phi}}^{{\rm s}}({\bf x}) .     
\end{equation}
Evaluation of this expression by using the charged partner (Eq.~(\ref{za3})) of
the Dirac monopole field yields
\begin{equation}
  \label{pm7a}
\delta m^{2}_{3} = \delta m^{2} .     
\end{equation}
Equation (\ref{pm5}) displays the resistance of the system to proliferate
production of monopoles. Furthermore since $\delta m^{2}$ decreases with
decreasing ``distance'' between monopoles, coupling to quantum fluctuations
effectively induces an attractive interaction between monopoles and
antimonopoles which, if strong enough, leads to monopole-antimonopole
annihilation and ultimately to a vanishing $\delta m^{2}$.  Also note the
confining nature of the interaction between magnetic monopoles of opposite
charge which is due to the standard Meissner effect with the field
$\hat{{\bf \Phi}}$ playing the role of the Higgs field. If still valid in
the confining phase, this perturbative treatment would point to a gluon mass
which, with decreasing temperature, increases with the monopole density
(cf.~Eq.~(\ref{pm11})). Without specifying the distribution of the monopoles, a
reliable estimate of the gluon mass is not possible. At high temperatures, as
argued above, we might expect the dynamics to favor monopoles of opposite
charge to be bound in dipoles. If the extension of the dipoles is much smaller
than their separation, the expression (\ref{pm5}) can be simplified and yields
a gluon mass
\begin{equation}
  \label{pm5a}
  \delta m^{2} = n_{{\rm M}} d
\end{equation}
determined by the monopole density and the average dipole size $d$. In such a
description of the high temperature phase as a gas of strongly correlated
magnetic monopoles, the appearance of a magnetic gluon mass is rather natural.
Temperature or coupling constant dependence of such a magnetic mass are
implicitly determined by monopole density and dipole size and cannot be
obtained by these simple arguments.
 
The picture of the monopole dynamics which we have developed is of some
relevance for the interpretation of results obtained in QCD lattice
calculations. The axial gauge discussed here is closely related to the Polyakov
gauge employed in lattice calculations. If the full local Abelian gauge
symmetry still present in the Polyakov gauge is fixed by eliminating
$A_{0}^{3}(x)$ up to the zero modes $a_{0}^{3}({\bf x})$, the axial gauge
representation is obtained. The properties of the monopoles are identical in
these two gauges. When comparing our qualitative arguments with numerical
results \cite{KRSW87} we find support for persistence of a monopole condensate
beyond the deconfinement transition. It is not known, though suggested by
calculations for $2+1$-dimensional QCD \cite{TRPW95}, whether the strong
correlations between poles and antipoles as indicated by our arguments indeed
exist in the high temperature phase nor has a relation between magnetic mass
and monopole properties like Eq.~(\ref{pm5a}) been established numerically.
Such results would be helpful for clarifying non-perturbative properties of the
high temperature phase. Lattice calculations exhibit strong dependence in the
monopole dynamics on the particular gauge choice. In the maximal Abelian gauge,
a significant monopole density is present only in the confined phase
\cite{KLSW87,SUZU93,POLI97}. This qualitative difference between the axial or
Polyakov gauge, on the one hand, and the maximal Abelian gauge on the other may
be due to the very different properties of the gauge condition in the weak
coupling limit. The differential form of the maximal Abelian gauge condition
\cite{THOO81,KRSW87},
\begin{equation}
  \label{pm6a}
  (\partial_{\mu} + ig A_{\mu}^{3})\Phi_{\mu} = 0 ,
\end{equation}
reduces for weak coupling to the standard Lorentz gauge. This gauge is (up to
overall zero modes) well defined. In the absence of gauge ambiguities no
monopoles are present. On the other hand, in the same weak coupling limit, the
Polyakov loop approaches the north pole
\begin{equation}
  \label{pm6b}
  P({\bf x}) \rightarrow
  \openone \quad \text{for} \quad g\rightarrow 0 ,
\end{equation}
i.e., space gets filled with static monopoles in axial or Polyakov gauge. This
suggests the possibility of a unified picture of monopole dynamics for the
confined phase only and monopole condensation as a signature of confinement
only in gauges which become well defined in the weak coupling limit.

In summary, we have identified in this qualitative discussion of the physics of
axial gauge monopoles two conflicting tendencies of monopole dynamics. On the
one side, entropy arguments naturally favor production of monopoles. Our
discussion moreover suggests an unusual increase in entropy with decreasing
temperature. On the other side, monopole interactions mediated by coupling to
quantum fluctuations induce attractive monopole-antimonopole interactions which
ultimately will induce annihilation of monopole-antimonopole pairs. These two
conflicting tendencies suggest an interpretation of the confined phase as a
phase with a finite and at $T=0$ divergent density of weakly correlated
monopoles, while in the deconfined phase, monopoles are bound into dipoles. The
confinement-deconfinement transition is thus similar to the phase transition in
the two-dimensional XY model which occurs by vortex unbinding, however with high
and low temperature phases interchanged. Unlike in standard systems, in QCD the
entropy associated with monopoles increases with decreasing temperature which
might be ultimately the source of the unusual phenomenon of spontaneous
symmetry breakdown in the high temperature phase.

\acknowledgments
This work has been supported by the Bundesministerium f\"ur Bildung,
Wissenschaft, Forschung und Technologie and the Monbusho International
Scientific Research Program (No.~06044053). We would like to thank M.~Thies for
his collaboration in the early stage of this work, U.-J.~Wiese and K.~Yazaki
for fruitful discussions. We express our gratitude to V.~Eletsky,
A.~Kalloniatis and L.~v.~Smekal for careful reading of the manuscript and
valuable comments and suggestions concerning both physics and presentation.
F.L. thanks K.~Yazaki for the hospitality extended to him at the University of
Tokyo where part of this work was done.

\end{document}